\documentclass{article}

\title{Is Cosine-Similarity of Embeddings Really About Similarity?}
\author{Harald Steck  \\
        hsteck@netflix.com\\
	Netflix Inc.  \\
        Los Gatos, CA, USA\\
	\and 
	Chaitanya Ekanadham\\
        cekanadham@netflix.com\\
        Netflix Inc.\\
        Los Angeles, CA, USA
        \and
        Nathan Kallus\\
        nkallus@netflix.com\\
        Netflix Inc. \& Cornell University\\
        New York, NY, USA
}

\usepackage{graphicx}
\usepackage{amsfonts}

\newcommand{\diagMat}{{\rm dMat}}

\newcommand{\RR}{\mathbb{R}}

\begin{document}

\maketitle

\begin{abstract}
  Cosine-similarity is the cosine of the angle between two vectors, or equivalently the dot product between their normalizations. A popular application is to quantify semantic similarity between high-dimensional objects by applying cosine-similarity to a learned low-dimensional feature embedding. This can work better but sometimes also worse than the unnormalized dot-product between embedded vectors in practice. To gain insight into this empirical observation, we study embeddings derived from regularized linear models, where closed-form solutions facilitate analytical insights. We derive analytically how cosine-similarity can yield arbitrary and therefore meaningless `similarities.' For some linear models the similarities are not even unique, while for others they are implicitly controlled by the regularization. We discuss implications beyond linear models: a combination of different regularizations are employed when learning deep models; these have implicit and unintended effects when taking cosine-similarities of the resulting embeddings, rendering results opaque and possibly arbitrary. Based on these insights, we caution against blindly using cosine-similarity and outline alternatives.
\end{abstract}

\section{Introduction}

Discrete entities are often embedded via a learned mapping to dense real-valued vectors in a variety of domains. For instance, words are embedded based on their surrounding context in a large language model (LLM), while recommender systems often learn an embedding of items (and users) based on how they are consumed by users. The benefits of such embeddings are manifold. In particular, they can be used directly as (frozen or fine-tuned) inputs to other models, and/or they can provide a data-driven notion of (semantic) similarity between entities that were previously atomic and discrete.

While \emph{similarity} in 'cosine similarity' refers to the fact that \emph{larger} values (as opposed to \emph{smaller} values in \emph{distance} metrics) indicate closer proximity, it has, however, also become a very popular measure of \emph{semantic} similarity between the entities of interest, the motivation being that the norm of the learned embedding-vectors is not as important as the directional alignment between the embedding-vectors. While there are countless papers that report the successful use of cosine similarity in practical applications, it was, however, also found to not work as well as other approaches, like the (unnormalized) dot-product between the learned embeddings, e.g., see  \cite{karpukhin20,khattab20,zhou22}.

In this paper, we try to shed light on these inconsistent empirical observations. We show that \emph{cosine similarity} of the learned embeddings can in fact yield arbitrary results. We find that the underlying reason is not cosine similarity itself, but the fact that the learned embeddings have a degree of freedom that can render arbitrary cosine-similarities even though their (unnormalized) dot-products are well-defined and unique. To obtain insights that hold more generally, we derive analytical solutions, which is possible for linear Matrix Factorization (MF) models--this is outlined in detail in the next Section.  In Section \ref{sec_sol}, we propose possible remedies. The experiments in 
 Section \ref{sec_exp} illustrate our findings derived in this paper.

%%%%%%%%%%%%%%%%%%%%%%%%%%%%%%%%%%%%%%%%%%%%%%%%%%%%%%%%%%%%%%%%%%%%%%%%%%%%%%
%%%%%%%%%%%%%%%%%%%%%%%%%%%%%%%%%%%%%%%%%%%%%%%%%%%%%%%%%%%%%%%%%%%%%%%%%%%%%%
\section{Matrix Factorization Models} \label{sec_mf}
In this paper, we focus on linear models as they allow for closed-form solutions, and hence a theoretical understanding of the limitations of the cosine-similarity metric applied to learned embeddings. We are given a matrix $X\in \RR^{n\times p}$ containing $n$ data points and $p$ features (e.g.,  users and items, respectively, in case of recommender systems). The goal in matrix-factorization (MF) models, or equivalently in linear autoencoders,  is to estimate a low-rank matrix $AB^\top \in \RR^{p \times p}$, where $A,B \in \RR^{p\times k}$ with $k \le p$, such that the product $XAB^T$ is a good approximation of $X$:\footnote{Note that we omitted bias-terms (constant offsets) here for clarity of notation--they can simply be introduced in the preprocessing step by subtracting them from each column or row of $X$. Given that such bias terms can reduce the popularity-bias of the learned embeddings to some degree, they can have some impact regarding the learned similarities, but it is ultimately limited.}
$ X \approx XAB^\top $.
When the given $X$ is a user-item matrix, the rows $\vec{b_i}$ of $B$ are typically referred to as the ($k$-dimensional) item-embeddings, while the rows of $XA$, denoted by $\vec{x_u}\cdot A$, can be interpreted as the user-embeddings, where the embedding of user $u$ is the sum of the item-embeddings $\vec{a_j}$ that the user has consumed.

Note that this model is defined in terms of the (unnormalized) dot-product between the user and item embeddings  $(XAB^\top)_{u,i}=\langle\vec{x_u}\cdot A, \vec{b_i}\rangle$. Nevertheless, once the embeddings have  been learned, it is common practice to also consider their cosine-similarity,  between two items $cosSim(\vec{b_i}, \vec{b_{i'}})$, two users $cosSim(\vec{x_u}\cdot A, \vec{x_{u'}}\cdot A)$, or a user and an item $cosSim(\vec{x_u}\cdot A, \vec{b_i})$. In the following, we show that this can lead to arbitrary results, and they may not even be unique.

\subsection{Training}

A key factor affecting the utility of cosine-similarity metric is the regularization employed when learning the embeddings in $A, B$, as outlined in the following. Consider the following two, commonly used, regularization schemes (which both have closed-form solutions, see Sections \ref{sec_obj1} and \ref{sec_obj2}:
\begin{eqnarray}
 && \min_{A,B} ||X-XAB^\top||_F^2 + \lambda ||AB^\top||_F^2  \label{eq_asym_mf}\\
 && \min_{A,B} ||X-XAB^\top||_F^2 + \lambda (||XA||_F^2 + ||B||_F^2) \label{eq_mf}
\end{eqnarray}
  The two training objectives obviously differ in their L2-norm regularization:
\begin{itemize}
\item In the first objective,  $||AB^\top||_F^2$ applies to their product.  In linear models,  this kind of L2-norm regularization can be shown to be equivalent to learning with denoising, i.e., drop-out in the input layer, e.g., see \cite{steck20b}. Moreover, the resulting prediction accuracy on held-out test-data was experimentally found to be superior to the one of the second objective \cite{jin21}. Not only in MF models, but also in deep learning it is often observed that denoising or drop-out (this objective) leads to better results on held-out test-data than weight decay (second objective) does.

  \item The second objective is equivalent to the usual matrix factorization objective $ \min_W ||X-PQ^\top||_F^2 + \lambda (||P||_F^2 + ||Q||_F^2)$, where $X$ is factorized as $PQ^\top$, and $P=XA$ and $Q=B$. This equivalence is outlined, e.g., in  \cite{jin21}. Here, the key is that each matrix $P$ and $Q$ is regularized separately, similar to weight decay in deep learning. 
\end{itemize}
If $\hat{A}$ and $\hat{B}$ are solutions to either objective, it is well known that then also $\hat{A}R$ and $\hat{B}R$ with an arbitrary rotation matrix $R\in\RR^{k \times k}$, are solutions as well. While cosine similarity is invariant under such rotations $R$, one of the  key insights in this paper is that the first (but not the second) objective is also invariant to rescalings of the columns of $A$ and $B$ (i.e., the different latent dimensions of the embeddings): if $\hat{A}\hat{B}^\top$ is a solution of the first objective, so is  $\hat{A}DD^{-1}\hat{B}^\top$  where $D \in \RR^{k \times k}$ is an arbitrary diagonal matrix. We can hence define a new solution (as a function of $D$) as follows:
\begin{eqnarray}
  \hat{A}^{(D)} &:=& \hat{A}D 
  \,\,\,\,\,\,\,\,\,\,\,\,\,\,\,\,\,\,\,\,\,\,
  {\rm and } \nonumber\\
  \hat{B}^{(D)} &:=& \hat{B}D^{-1}.
\end{eqnarray}
In turn, this diagonal matrix $D$ affects the normalization of the learned user and item embeddings (i.e., rows):
\begin{eqnarray}
  (X\hat{A}^{(D)})_{\rm (normalized)} &=& \Omega_A  X \hat{A}^{(D)}  = \Omega_A X \hat{A}D
    \,\,\,\,\,
  {\rm and }   \nonumber \\
  \,\,\,\,\,\,\,\,\,\,\,\,\,\,\,\,\,\,\,\,\,\,
  \hat{B}^{(D)}_{\rm (normalized)} &=& \Omega_B  \hat{B}^{(D)}  = \Omega_B \hat{B}D^{-1},
\end{eqnarray}
where $ \Omega_A$ and $\Omega_B$ are appropriate diagonal matrices to normalize each learned embedding (row) to unit Euclidean norm. Note that in general the matrices do not commute, and hence a different choice for $D$ \emph{cannot} (exactly) be compensated by the normalizing matrices $\Omega_A$ and $\Omega_B$. As they depend on $D$, we make this explicit by $\Omega_A(D)$ and $\Omega_B(D)$. \emph{Hence, also the cosine similarities of the embeddings depend on this arbitrary matrix $D$.} 

As one may consider the cosine-similarity between two items, two users, or a user and an item, the three combinations read
\begin{itemize}

\item item -- item:
$$
 {\rm cosSim}\left( \hat{B}^{(D)},  \hat{B}^{(D)} \right ) =
  %\hat{B}^{(D)}_{\rm (normalized)} \hat{B}^{(D)}_{\rm (normalized)^\top}
   \Omega_B(D)   \cdot  \hat{B} \cdot D^{-2}  \cdot \hat{B}^{\top}  \cdot  \Omega_B(D)
$$

\item  user -- user:
$$
   {\rm cosSim}\left( X\hat{A}^{(D)},  X\hat{A}^{(D)} \right ) 
  = \Omega_A(D)  \cdot   X\hat{A}  \cdot D^2  \cdot  (X\hat{A})^{\top}  \cdot  \Omega_A(D)
$$

\item user -- item:   
$$
  {\rm cosSim}\left( X\hat{A}^{(D)},  \hat{B}^{(D)} \right ) =
  %\hat{A}^{(D)}_{\rm (normalized)} \hat{B}^{(D)}_{\rm (normalized)\top}
   \Omega_A(D)  \cdot  X\hat{A}  \cdot   \hat{B}^{\top}  \cdot  \Omega_B(D)
$$

\end{itemize}
It is apparent that the cosine-similarity in all three combinations depends on the arbitrary diagonal matrix $D$: 
while they all indirectly depend on $D$ due to its effect on the  normalizing matrices $\Omega_A(D)$ and $\Omega_B(D)$, note that the (particularly popular) item-item cosine-similarity (first line) in addition depends directly on $D$ (and so does the user-user cosine-similarity, see second item).

\subsection{Details on First Objective (Eq. \ref{eq_asym_mf})}  
\label{sec_obj1}
%%%%%%%%%%%%%%%%%%%%%%%%%%%%%%%%%%%%%%%%%%%%%%%%%%%%%%%%%%%%%%%%%%%%%%%%%%%%%%
The closed-form solution of the training objective in Eq. \ref{eq_asym_mf} was derived in \cite{jin21} and reads
$
\hat{A}_{(1)}\hat{B}_{(1)}^\top = V_k \cdot \diagMat(..., \frac{1}{1+\lambda / \sigma^2_i} ,...  )_k \cdot V_k^\top   
$,
where $X =: U\Sigma V^\top$ is the singular value decomposition (SVD)  of the given data matrix $X$, where $\Sigma = \diagMat(..., \sigma_i  ,...)$ denotes the diagonal matrix of singular values, while $U,V$ contain the left and right singular vectors, respectively.  Regarding the $k$ largest eigenvalues $\sigma_i$, we denote the truncated matrices of rank $k$ as  $U_k, V_k$ and $(...)_k$. We may define\footnote{As $D$ is arbitrary, we chose to assign $\diagMat(..., \frac{1}{1+\lambda / \sigma^2_i} ,...  )_k^{\frac{1}{2}} $ to each of $\hat{A}, \hat{B}$ without loss of generality.}
\begin{equation}
  \hat{A}_{(1)}=\hat{B}_{(1)} :=  V_k    \cdot \diagMat(..., \frac{1}{1+\lambda / \sigma^2_i} ,...  )_k^{\frac{1}{2}}.
  \label{eq_v1_a}
  \end{equation}
The arbitrariness of cosine-similarity becomes especially striking here when we consider the special case of a full-rank MF model, i.e., when $k=p$. This is illustrated by the following two cases:
\begin{itemize}

\item if we choose $D= \diagMat(..., \frac{1}{1+\lambda / \sigma^2_i} ,...  )^{\frac{1}{2}}$, then we have $\hat{A}_{(1)}^{(D)} = \hat{A}_{(1)} \cdot D   = V\cdot\diagMat(..., \frac{1}{1+\lambda / \sigma^2_i} ,...  )$ and   $\hat{B}_{(1)}^{(D)} = \hat{B}_{(1)} \cdot D^{-1} = V$. Given that the full-rank matrix of singular vectors $V$ is already normalized (regarding both columns and rows), the normalization $\Omega_B=I$ hence equals the identity matrix $I$. We thus obtain regarding the  item-item cosine-similarities:  
$${\rm cosSim}\left( \hat{B}_{(1)}^{(D)},  \hat{B}_{(1)}^{(D)} \right ) =  VV^\top=I,$$ 
which is quite a bizarre  result, as it says that the  cosine-similarity between any pair of (different) item-embeddings is zero, i.e., an item is only similar to itself, but not to any other item!

Another remarkable result is obtained for the user-item cosine-similarity:
\begin{eqnarray*}
{\rm cosSim}\left( X\hat{A}_{(1)}^{(D)},  \hat{B}_{(1)}^{(D)} \right ) &=& \Omega_A   \cdot X \cdot V \cdot  \diagMat(..., \frac{1}{1+\lambda / \sigma^2_i} ,...  ) \cdot  V^{\top}  \\
&=&    \Omega_A \cdot X \cdot \hat{A}_{(1)}\hat{B}_{(1)}^\top, 
\end{eqnarray*}
   as the
 only difference to the (unnormalized) dot-product is due to the matrix $\Omega_A$, which normalizes the rows---hence, when we consider the ranking of the items for a given user based on the predicted scores, cosine-similarity and (unnormalized) dot-product result in exactly the same ranking of the items as the row-normalization is only an irrelevant constant in this case.

\item if we choose $D= \diagMat(..., \frac{1}{1+\lambda / \sigma^2_i} ,...  )^{-\frac{1}{2}}$, then we have analogously to the previous case:   $\hat{B}_{(1)}^{(D)} = V\cdot \diagMat(..., \frac{1}{1+\lambda / \sigma^2_i} ,...  ) $, 
and $\hat{A}_{(1)}^{(D)} = V$ is orthonormal. We now obtain regarding the  user-user cosine-similarities: 
$$ {\rm cosSim}\left( X\hat{A}_{(1)}^{(D)},  X\hat{A}_{(1)}^{(D)} \right ) = \Omega_A  \cdot   X  \cdot X^\top \cdot \Omega_A ,$$
i.e., now the user-similarities are simply based on the raw data-matrix $X$, i.e., without any smoothing due to the learned embeddings. Concerning the user-item cosine-similarities, we now obtain
$$
{\rm cosSim}\left( X\hat{A}_{(1)}^{(D)},  \hat{B}_{(1)}^{(D)} \right ) = \Omega_A\cdot X\cdot \hat{A}_{(1)}\cdot \hat{B}_{(1)}^\top \cdot  \Omega_B , $$
  i.e., 
   now $\Omega_B$ normalizes the rows of $B$, which we did not have in the previous choice of $D$. 
   
   Similarly, the item-item cosine-similarities 
   $$
   {\rm cosSim}\left( \hat{B}_{(1)}^{(D)},  \hat{B}_{(1)}^{(D)} \right ) 
   =\Omega_B   \cdot V\cdot \diagMat(..., \frac{1}{1+\lambda / \sigma^2_i} ,...  )^2 \cdot  V^{\top}  \cdot  \Omega_B
   $$
   are very different from the bizarre result we obtained in the previous choice of $D$. 

\end{itemize}
Overall, these two cases show that different choices for $D$ result in different cosine-similarities, even though the learned model $\hat{A}_{(1)}^{(D)}\hat{B}_{(1)}^{(D)\top} = \hat{A}_{(1)}\hat{B}_{(1)}^\top$ is invariant to $D$. In other words, the results of cosine-similarity are arbitray and not unique for this model.

\subsection{Details on Second Objective (Eq. \ref{eq_mf})} 
\label{sec_obj2}
%%%%%%%%%%%%%%%%%%%%%%%%%%%%%%%%%%%%%%%%%%%%%%%%%%%%%%%%%%%%%%%%%%%%%%%%%%%%%%
The solution of the training objective in Eq. \ref{eq_mf} was derived in \cite{zheng18} and reads
\begin{eqnarray}
  \hat{A}_{(2)} &=& V_k \cdot \diagMat(...,\sqrt{\frac{1}{\sigma_i}\cdot (1-\frac{\lambda}{\sigma_i})_+}   ,...)_k 
   \,\,\,\,\,\,\,\,\,\,\,\,\,\,\,\,\,
  {\rm and } \nonumber\\
  \hat{B}_{(2)} &=& V_k \cdot \diagMat(...,\sqrt{\sigma_i\cdot (1-\frac{\lambda}{\sigma_i})_+}   ,...)_k 
  \label{eq_p}
\end{eqnarray}
where $(y)_+=\max(0,y)$, and again $X =: U\Sigma V^\top$ is the SVD of the training data $X$, and $\Sigma = \diagMat(..., \sigma_i  ,...)$.  
Note that, if we use the usual notation of MF where $P=XA$ and $Q=B$, we obtain
$
 \hat{P} = X\hat{A}_{(2)}  = U_k \cdot \diagMat(...,\sqrt{\sigma_i\cdot (1-\frac{\lambda}{\sigma_i})_+}   ,...)_k  
 $,
 where we can see that here the diagonal matrix $\diagMat(...,\sqrt{\sigma_i\cdot (1-\frac{\lambda}{\sigma_i})_+}   ,...)_k  $ is the same for the user-embeddings and the item-embeddings in Eq. \ref{eq_p}, as expected due to the symmetry in the L2-norm regularization $||P||_F^2+||Q||_F^2$ in the training objective in Eq. \ref{eq_mf}. 

 The key difference to the first training objective (see Eq. \ref{eq_asym_mf}) is that here  the  L2-norm regularization $||P||_F^2+||Q||_F^2$ is applied to each matrix individually, so that this solution is unique (up to irrelevant rotations, as mentioned above), i.e., in this case there is no way to introduce an arbitrary diagonal matrix $D$ into the solution of the second objective. Hence, the cosine-similarity applied to the learned embeddings of this MF variant yields unique results.

 While this solution is unique, it remains an open question if this unique diagonal matrix $\diagMat(...,\sqrt{\sigma_i\cdot (1-\frac{\lambda}{\sigma_i})_+}   ,...)_k  $ regarding the user and item embeddings yields the best possible semantic similarities in practice.
 If we believe, however, that this regularization makes the cosine-similarity useful concerning semantic similarity, we could compare the forms of the diagonal matrices in both variants, i.e., comparing Eq. \ref{eq_p} with Eq. \ref{eq_v1_a}  suggests that the arbitrary diagonal matrix $D$ in the first variant (see section above) analogously may be chosen as $D = \diagMat (..., \sqrt{1/\sigma_i}  ,...)_k$.

\section{Remedies and Alternatives to Cosine-Similarity}
\label{sec_sol}
As we showed analytically above, when a model is trained w.r.t. the dot-product, its effect on cosine-similarity can be opaque and sometimes not even unique. 
One solution obviously is to train the model w.r.t. to cosine similarity, which layer normalization \cite{ba16} may facilitate. Another approach is to avoid the embedding space, which caused the problems outlined above in the first place, and project it back into the original space, 
where the cosine-similarity can then be applied. For instance, using the models above, and given the raw data $X$, one may view $X\hat{A}\hat{B}^\top$ as its smoothed version, and the rows of  $X\hat{A}\hat{B}^\top$ as the users' embeddings in the original space, where cosine-similarity may then be applied.

Apart from that, it is also important to note that,  in cosine-similarity, normalization  is applied only \emph{after} the embeddings have been learned. This can noticeably reduce the resulting (semantic) similarities compared to applying some normalization, or reduction of popularity-bias, \emph{before} or \emph{during}  learning. This can be done in several ways. For instance, a default approach in statistics is to standardize the data $X$ (so that each column has zero mean and unit variance). Common approaches in deep learning include the use of negative sampling  or inverse propensity scaling (IPS) as to account for the different item popularities (and user activity-levels). For instance, in word2vec \cite{mikolov13b}, a matrix factorization model was trained by sampling  negatives  with a probability proportional to their frequency (popularity) in the training data taken to the power of $\beta =3/4$, which resulted in impressive word-similarities at that time.

\begin{figure}[t]
    \includegraphics[height=2.65cm]{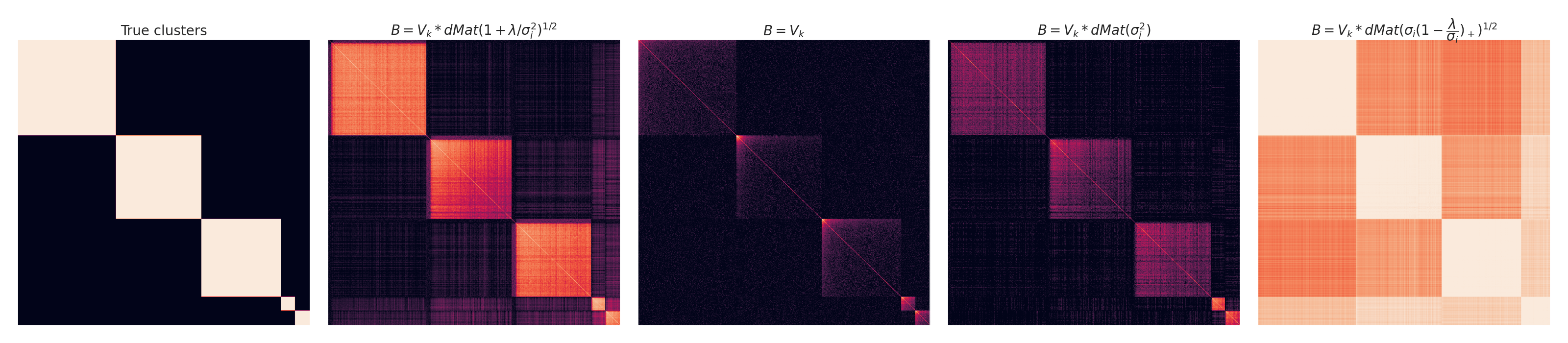}
    \caption{Illustration of the large variability of item-item cosine similarities $cosSim(B,B)$ on the same data due to different modeling choices. Left: ground-truth clusters (items are sorted by cluster assignment, and within each cluster by descending baseline popularity). After training w.r.t. Eq. \ref{eq_asym_mf}, which allows for arbitrary re-scaling of the singular vectors in $V_k$, the center three plots show three particular choices of re-scaling, as indicated above each plot. % with $B = V_k * dMat(...,\frac{1}{1 + \lambda / \sigma_i^2})^{\frac{1}{2}}$ (i.e. Eq.~\ref{eq_v1_a}). Middle: Eq. \ref{eq_asym_mf} with $B = V_k$. Middle right: Eq. \ref{eq_asym_mf} with $B = V_k * dMat(...,\sigma_i^2,...)$. 
    Right: based on (unique) $B$ obtained when training w.r.t. Eq. \ref{eq_mf}.}
    \label{fig_low_rank}
\end{figure}

\section{Experiments}
\label{sec_exp}
While we discussed the full-rank model above, as it was amenable to analytical insights, we now illustrate these findings experimentally for low-rank embeddings. We are not aware of a good metric for semantic similarity, which motivated us to conduct experiments on simulated data, so that the ground-truth semantic similarites are known. To this end, we simulated data where items are grouped into clusters, and users interact with items based on their cluster preferences. We then examined to what extent cosine similarities applied to learned embeddings can recover the item cluster structure.

In detail, we generated interactions between $n=20,000$ users and $p=1,000$ items that were randomly assigned to $C=5$ clusters with probabilities $p_c$ for $c=1,...,C$. Then we sampled the powerlaw-exponent for each cluster $c$, 
$\beta_c \sim {\rm Unif}(\beta_{min}^{(item)}, \beta_{max}^{(item)})$ where we chose $\beta_{min}^{(item)}=0.25$ and $\beta_{max}^{(item)}=1.5$, and then assigned a baseline popularity to each item $i$ according to the powerlaw $p_{i} = {\rm PowerLaw}(\beta_c)$. Then we generated the items that each user $u$ had interacted with: first, we randomly sampled user-cluster preferences $p_{uc}$, and then  computed the user-item probabilities: $p_{ui} = \frac{p_{u{c_i}}p_{i}}{\sum_{i}p_{u{c_i}}p_{i}}$. We sampled the number of items for this user, $k_u \sim \rm{PowerLaw}(\beta^{(user)})$, where we used $\beta^{(user)} = 0.5$, and then sampled  $k_u$ items (without replacement) using probabilities $p_{ui}$.

We then learned the matrices $A, B$ according to Eq. \ref{eq_asym_mf} and also Eq. \ref{eq_mf} (with $\lambda=10,000$ and $\lambda= 100$, respectively) from the simulated data. We used a low-rank constraint $k=50 \ll p=1,000$ to complement the analytical results for the full-rank case above.

Fig.~\ref{fig_low_rank} shows the "true" item-item-similarities as defined by the item clusters on the left hand side, while the remaining four plots show the item-item cosine similarities obtained for the following four scenarios:  after training w.r.t. Eq. \ref{eq_asym_mf}, which allows for arbitrary re-scaling of the singular vectors in $V_k$ (as outlined in  Section \ref{sec_obj1}), the center three cosine-similarities are obtained for three choices of re-scaling. The last plot in this row is obtained from training w.r.t. Eq. \ref{eq_mf}, which results in a unique solution for the cosine-similarities. Again, the main purpose here is to illustrate how vastly different the resulting cosine-similarities can be even for reasonable choices of re-scaling when training w.r.t. Eq. \ref{eq_asym_mf} (note that we did not use any extreme choice for the re-scaling here, like anti-correlated with the singular values, even though this would also be permitted), and also for the unique solution when training w.r.t. Eq. \ref{eq_mf}.

\section*{Conclusions}

It is common practice to use cosine-similarity between learned user and/or item embeddings as a measure of semantic similarity between these entities. We study cosine similarities in the context of linear matrix factorization models, which allow for analytical derivations, and show that cosine similarities are heavily dependent on the method and regularization technique, and in some cases can be rendered even meaningless. Our analytical derivations are complemented experimentally by qualitatively examining the output of these models applied simulated data where we have ground truth item-item similarity. Based on these insights, we caution against blindly using cosine-similarity, and proposed a couple of approaches to mitigate this issue. While this short paper is limited to linear models that allow for insights based on analytical derivations, we expect cosine-similarity of the learned embeddings in \emph{deep models } to be plagued by similar problems, if not larger ones, as  a combination of various regularization methods is typically applied there, and different layers in the model may be subject to different regularization---which implicitly determines a particular scaling (analogous to matrix $D$ above) of the different latent dimensions in the learned embeddings in each layer of the deep model, and hence its effect on the resulting cosine similarities may become even more opaque there. 

  \bibliographystyle{plain}
  \bibliography{mybib}
\end{document}